\documentstyle{article}
\def\m@th{\mathsurround=0pt}
\def\n@space{\nulldelimiterspace=0pt \m@th}
\def\biggg#1{{\mbox{$\left#1\vbox to 20.5pt{}\right.\n@space$}}}
\def\cd{\hbox{\kern-.02em\hbox{$\cdot$}\kern-.01em}}
\newcommand{\bfna}{{\mbox{\boldmath $\bf\nabla$}}}
\def\p{\partial}
\def\rot{\bfna\times}
\def\dv{\bfna\cd}
\def\const{{\rm const}}
\def\bq{\begin{equation}}
\def\eq{\end{equation}}
\def\eqn{$$}
\def\bqn{$$}
\def\bqy{\begin{eqnarray}}
\def\eqy{\end{eqnarray}}
\def\bqyn{\begin{eqnarray*}}
\def\eqyn{\end{eqnarray*}}
\def\bc{\begin{center}}
\def\ec{\end{center}}
\def\bfV{{\bf V}}
\def\bfB{{\bf B}}
\def\bfD{{\bf D}}
\def\bfF{{\bf F}}
\def\bfr{{\bf r}}
\def\bfv{{\bf v}}
\def\bfu{{\bf u}}
\def\bfw{{\bf w}}
\newcommand{\bfxi}{{\mbox{\boldmath $\bf\xi$}}}
\newcommand{\bfeta}{{\mbox{\boldmath $\bf\eta$}}}
\begin{document}
\title{\bf     Variational  principle for linear stability of
moving magnetized plasma.}
\author{Victor I. Ilgisonis\\
{\small email:~~vil@nfi.kiae.ru}\\
{\it{Russian Research Centre ``Kurchatov Institute''}}\\
{\it{123182 Moscow, Russia}}} \maketitle
\begin{abstract}
The variational principle for linear stability of
three-dimensional, inhomogenious, compressible, moving magnetized
plasma is suggested. The principle is ``softer'' (easier to be
satisfied) than all previously known variational stability
conditions. The key point of the analysis is a conservation in
variations of new integrals inherent in the linearized equation of
the motion that was not earlier discussed in the literature.
\end{abstract}
PACS{46.15.Cc, 52.30.Cv}


It is well known that stability of the static equilibrium of
magnetized plasma can be described by so-called ``energy
principle'' \cite{1}. The principle claims that if the second
variation of the potential energy, $W$, of the system
``plasma-magnetic field'' is positive definite near the
equilibrium point, then this point is stable. Sufficiency of the
claim follows from the Lyapunov stability theorem, and necessity
can also be proved \cite{1}. Note that the above mentioned second
variation of the potential energy corresponds exactly to the
potential energy of the linearized equation system.

      The main drawback of the principle is that there always
may be neutral perturbations which do not perturb
any physical quantity -- and, therefore, $W$ as well.
Thus, the second variation of the potential  energy can
be guaranteed to be only positive semi-definite. In other
words, using the energy principle \cite{1}, one can talk about
spectral stability only, namely, about presence or absence of
imaginary frequencies in the spectrum of the linearized force
operator (nonlinear stability needs an
analysis of neutral perturbations -- see, e.g., \cite{my14}).
Continuing this logic, we restrict ourselves with
linearized  equations.

     The attempt of using the similar approach to investigate
stability of moving plasma performed by Frieman
and Rotenberg \cite{FR} was not so lucky, although the energy
principle was formally obtained. Their result can be
briefly described as follows. Consider the linearized
equation of motion for plasma displacement $\bfxi$ in the frame
of ideal one-fluid magnetohydrodynamics,
\bq
\label{1}
\rho\ddot \bfxi+2\rho(\bfV\cd\bfna)\dot \bfxi-\bfF (\bfxi)=0\ ,
\eq
where the linearized force operator,
\bqyn
\bfF(\bfxi)&=&-\delta\rho(\bfV\cd\bfna)\bfV-\rho(\delta\bfV\cd\bfna)\bfV-
\rho(\bfV\cd\bfna)\delta\bfV\\
&-&\bfna\delta p+(\rot\delta\bfB)\times\bfB
+(\rot\bfB)\times\delta\bfB\ , \eqyn is combined of usual
perturbed quantities, \bqn
\delta\rho=-\dv(\rho\bfxi),~~~\delta\bfV=(\bfV\cd\bfna)\bfxi
-(\bfxi\cd\bfna)\bfV\ , \eqn \bqn \delta p=-\bfxi\cd\bfna p-\gamma
p\,\dv\bfxi,~~~\delta\bfB=\rot(\bfxi\times\bfB) \eqn
(note that $\delta\bfV$ denotes here only the part of full Eulerian
velocity perturbation -- the part, which survives even for
time-independent displacements, $\bfxi$).
Stationary
plasma density, $\rho$, velocity, $\bfV$, pressure, $p$, and
magnetic field, $\bfB$, satisfy the following equilibrium
conditions: \bqyn
&&\rho(\bfV\cd\bfna)\bfV+\bfna p=(\rot\bfB)\times\bfB\ ,\nonumber\\
&&\dv(\rho\bfV)=0\ ,\\
&&\bfV\cd\bfna p+\gamma p\,\dv\bfV=0\ ,\\
&&\rot(\bfV\times\bfB)=0 \ . \eqyn Dot means a partial
time-derivative, $\gamma$ means the adiabatic exponent. Force
operator is proved to be self-adjoint in the following sense, \bqn
\int\bfeta\cd\bfF(\bfxi)\, d^3 r=\int\bfxi\cd\bfF(\bfeta)\, d^3 r\
, \eqn while the second term in (\ref{1}) is obviously
antisymmetric: \bqn \int\bfeta\cd\rho(\bfV\cd\bfna)\bfxi \, d^3
r=-\int\bfxi\cd\rho(\bfV\cd\bfna)\bfeta \, d^3 r\ . \eqn
Multiplying Eq.~(1) by $\dot\bfxi$ and integrating over the whole
space, we found the energy conservation in the form $\dot E=0$,
where \bq \label{E1}
E(t)=\int\left(\rho\frac{\dot\bfxi^2}{2}-\frac{\bfxi\cd
\bfF(\bfxi)}{2}\right)\, d^3 r\ . \eq Minimizing $E$ over $\dot
\bfxi$, we approach to the energy principle by Frieman-Rotenberg,
\bq \label{st1} -\int\bfxi\cd\bfF(\bfxi)\, d^3 r\ge0\ . \eq
Contrary to the static case $(\bfV=0)$, in which condition
(\ref{st1}) appears to be both sufficient and necessary for linear
stability, in the case of $\bfV\ne 0$, condition (\ref{st1}) is
normally too strong, and never can be satisfied except of
field-aligned flows $(\bfV\sim\bfB)$\cite{FR} or of those which
may be reduced to the field-aligned flows (see, e.g., \cite{ILVL}).

Energy principle (\ref{st1}) may be improved by
use of the Arnold conjecture \cite{5}-\cite{6}, following
which we have to add to the energy (\ref{E1}) the set of
other known integrals of the motion. Speaking in other words,
variables $\dot \bfxi$ and $\bfxi$ in (\ref{E1}) are not
absolutely independent but subject to the constraints resulting
from conservation of other integrals of the motion.

Such an improved principle was derived by Ilgisonis and Pastukhov
\cite{7}, then it was verified by
Hameiri \cite{8}. It was also re-obtained \cite{9} with help of
Pfirsch \& Morrison's method \cite{10} of dynamically accessed
perturbations.
That stability condition is currently the best among
the known ones, although it is still not
appropriate for
arbitrary stationary plasma flow.

For the linearized equation (\ref{1}), the
Ilgisonis \& Pastukhov extra invariants can be written in terms of
neutral perturbation $\bfxi_N$: \bqn
\bfF(\bfxi_N)=0,~~~\p_t\bfxi_N=0\ . \eqn Multiplying Eq.~(\ref{1})
by $\bfxi_N$ and integrating again over the space, we have $\dot
I=0$, where \bq \label{I1} I=\int(\rho\dot
\bfxi\cd\bfxi_N+2\rho\bfxi_N(\bfV\cd\bfna)\bfxi)\, d^3 r\ . \eq
For the system with nested set of magnetic surfaces,
$\psi(\bfr)=\const$, $\bfxi_N$ may be generally represented as \bq
\label{xiN} \bfxi_N=\lambda_u(\psi)\bfu+\lambda_v(\psi)\bfv\ , \eq
where $\bfu=\bfB/\rho, \bfv=\bfD/\rho$, and $\bfD$ is a
divergence-free frozen-in-plasma vector, tangential to the same
magnetic surfaces, $\psi(\bfr)=\const$, but different from $\bfB$,
\bq \label{BD} \bfB\times\bfD=\rho\bfna\psi \eq -- see \cite{7,8}
for explanations of how $\bfD$ can be built-up. For $\lambda_v=0$
in (\ref{xiN}), conservation of $I$ corresponds to the
cross-helicity invariance, which is well known contrary to more
general quantity $I$. Note that taking into account (\ref{xiN}),
(\ref{BD}), the second term under the integral in Eq.~(\ref{I1}) can
also be written as
$$-2\bfxi\cd\rho(\bfV\cd\bfna)\bfxi_N\ ,$$ or as
\bqn
\rho\bfxi_N\cd(\bfV\cd\bfna)\bfxi+\rho\bfV\cd(\bfxi_N\cd\bfna)\bfxi\ ,
\eqn
for absolutely arbitrary functions $\lambda_{u,v}(\psi).$

It is very important that Eq.~(\ref{1}) allows for some
extra set of invariants different from (\ref{E1}) and (\ref{I1}).
Indeed, differentiating Eq.~(\ref{1}) with respect to time, then
multiplying it by $\ddot \bfxi$ and integrating, we found -- like in
the case of energy but using once again original equation of
motion (\ref{1}) --
that the following quantity,
\bq
\label{E2}
E_2=\frac{1}{2}\int\left\{ \frac{1}{\rho}\left(F(\bfxi)-
2\rho\bfV\cd\bfna\dot \bfxi\right)^2-
\dot \bfxi\cd\bfF(\dot\bfxi)\right\} \, d^3 r\ ,
\eq
is conserved. This invariant is exact for linearized dynamics (\ref{1}),
and cannot be reduced to the conservation
of energy (\ref{E1}). In principle, we may continue the
procedure and get in the same manner an infinite set of similar
invariants. However, to investigate a stability, it might be
sufficient to involve into our analysis only finite number of
the invariants. Here we show that taking into account even one
of them, $E_2$, it appears to be possible to improve the
stability condition significantly.

Note that being varied separately (i.e., when $\dot \bfxi$ is
independent), invariant (\ref{E2}) results in the same
Frieman-Rotenberg condition of semi-positive definiteness of
quadratic form (\ref{st1}) based on $\bfF$-operator. To improve
stability condition, we will
consider the functional \bq \label{U} U=E+\mu E_2-I \eq to be
varied over $\bfxi$ and $\dot \bfxi$, subject to the independent
conservation of the integrals, $E_2$ and $I$. Explicitly, \bqy
\label{Ue} U=\int&&\bigg\{\frac{\rho\dot
\bfxi^2}{2}-\frac{\bfxi\cd\bfF(\bfxi)}{2}
+\frac{\mu}{2\rho}\left(\bfF(\bfxi)-2\rho(\bfV\cd\bfna)\dot
\bfxi\right)^2-
\frac{\mu}{2}\dot \bfxi\cd\bfF(\dot \bfxi)\nonumber\\
&&+(2\rho\bfxi\cd(\bfV\cd\bfna)-\rho\dot\bfxi\cd)(\lambda_u\bfu
+\lambda_v\bfv)\bigg\} \, d^3 r\ , \eqy where the constant, $\mu$,
and 1-D functions, $\lambda_{u,v}(\psi)$, play roles of the
Lagrangian multipliers; we have to choose them to provide the
integrals $E_2,~I$ be equal to their equilibrium values, i.e.,
to zero.

Functional $U$ is minimized by $\dot \bfxi$: \bq \label{xim} \dot
\bfxi=\underbrace{\lambda_u\bfu+\lambda_v\bfv}_{\bfxi_N}
+\mu\frac{\bfF(\dot \bfxi)}{\rho}
+2\mu\bfV\cd\bfna\bigg(2(\bfV\cd\bfna)\dot
\bfxi-\frac{\bfF(\bfxi)}{\rho}\bigg)\ . \eq Putting $\mu\to 0$ in
Eqs.~(\ref{Ue}), (\ref{xim}), we approach to the
Ilgisonis-Pastukhov-Hameiri condition \cite{7,8}. Indeed, we have
at $\mu\to 0$: \bqn \dot \bfxi\to\bfxi_N,~~U\to U_{IPH}=\int \,
d^3 r(\frac{\rho \bfxi_N^2}{2}-\frac{\bfxi\cd\bfF(\bfxi)}{2})\ ,
\eqn where $\bfxi_N$ satisfies (\ref{xiN}), and $\lambda_{u,v}$:
\bqn \int(\rho\bfxi_N^2-2\rho\bfxi\cd(\bfV\cd\bfna)\bfxi_N) \, d^3
r=0\ . \eqn In this limit, we have not really used the condition
of $E_2$-conservation. Note that $U_{IPH}\ge -\int
{\bfxi\cd\bfF(\bfxi)}\, d^3 r/2$, and, therefore, the
Ilgisonis-Pastukhov-Hameiri condition, $U_{IPH}\ge 0,$ is "softer"
than the condition (\ref{st1}) by Frieman and Rotenberg. However,
it is still not appropriate for arbitrary flow, in which $\bfV$ is
not parallel to $\bfB$. The sign-indefinite term,
${\bfxi\cd\bfF(\bfxi)}\ ,$ contains the high-order
$\bfxi$-derivatives and, therefore, can always prevail on the
positive term $\rho\bfxi_N^2$ (see \cite{8} for more details).

Now let us account for small but finite $\mu$. Solving
Eq.~(\ref{xim}) by iterations in $\mu$, we found \bq \label{xid}
\dot \bfxi\approx\bfxi_N-2\mu(\bfV\cd\bfna)\ddot \bfxi_0\ , \eq
where we used the notation \bq \label{ddxi0} \ddot
\bfxi_0=\bfF(\bfxi)/\rho-2(\bfV\cd\bfna)\bfxi_N\ . \eq Stability
condition is expressed again by functional $U$ depending on $\bfxi$:
\bq \label{st2}
U=\int\bigg\{\frac{\rho}{2}(\bfxi_N-2\mu(\bfV\cd\bfna)\ddot
\bfxi_0)^2- \frac{1}{2}\bfxi\cd\bfF(\bfxi)\bigg\} \, d^3 r\ge0\ .
\eq Here $\bfxi_N$ and $\ddot \bfxi_0$ are defined by
Eqs.~(\ref{xiN}), (\ref{ddxi0}), and also depend on $\bfxi$.

Lagrangian multipliers have to be found by substituting
Eq.~(\ref{xid}) into the conditions \bqn E_2, I~(\dot
\bfxi,\bfxi)\biggg|_{
\dot\bfxi=\bfxi_N- 2\mu(\bfV\cd\bfna)\ddot \bfxi_0}\approx 0\ .
\eqn They are: \bq \label{mum} \mu=\frac{1}{8}\frac{\int\rho\ddot
\bfxi_0^2 ~\, d^3 r}{\int\rho((\bfV\cd\bfna)\ddot \bfxi_0)^2 \,
d^3 r}\ , \eq \hskip 0.4cm \bq \label{lam}
\lambda_u=\frac{A_vD_u-A_0D_v}{A_uA_v-A_0^2}\ ,~~~
\lambda_v=\frac{A_uD_v-A_0D_u}{A_uA_v-A_0^2}\ . \eq Here \bqn
A_{w=u,v}=<4\mu\rho((\bfV\cd\bfna)\bfw)^2-\rho\bfw^2>\ , \eqn \bqn
A_0=<4\mu\rho((\bfV\cd\bfna)\bfu)\cd((\bfV\cd\bfna)\bfv)-\rho\bfu\cd\bfv>\
, \eqn \bqn
D_{w=u,v}=<2(\mu\bfF(\bfxi)-\rho\bfxi)\cd(\bfV\cd\bfna)\bfw>\ ,
\eqn and angular brackets mean the averaging over magnetic surface.

Note that the left-hand-side of the condition (\ref{st2})
contains the high-order derivatives of $\bfxi$ in the first
(non-negative) term, hence, the second (sign-indefinite) term is no
more critical. It is the main advantage of the condition
(\ref{st2}) with respect to previous one, $U_{IPH}\ge 0$,
that it has a sense for
arbitrary (not only field-aligned) flow and, therefore, may
have a practical merit.

This work was partially supported by the Human Capital
Foundation Grant No. 41.

\end{document}